# The Product Space Conditions the Development of Nations

C.A. Hidalgo.* ✝ (1), B. Klinger*(2), A.-L. Barabási(1), R. Hausmann(2).

(1) Center for Complex Network Research and Department of Physics, University of Notre Dame, Notre Dame, IN. 46556.
(2) Center for International Development, Kennedy School of Government, Harvard University, Cambridge, MA. 02139.

✝ To whom correspondence should be addressed: chidalgo@nd.edu.
*These authors contributed equally to this work.

*Abstract*
*Economies grow by upgrading the type of products they produce and export. The technology, capital, institutions and skills needed to make such new products are more easily adapted from some products than others. We study the network of relatedness between products, or 'product space', finding that most upscale products are located in a densely connected core while lower income products occupy a less connected periphery. We show that countries tend to move to goods close to those they are currently specialized in, allowing nations located in more connected parts of the product space to upgrade their exports basket more quickly. Most countries can reach the core only if they "jump" over empirically infrequent distances in the product space. This may help explain why poor countries have trouble developing more competitive exports, failing to converge to the income levels of rich countries.*

Does the type of product a country exports matter for subsequent economic performance? In David Ricardo's words, does it matter if Britain specializes in cloth and Portugal in wine? The fathers of development economics held that it does, suggesting that industrialization creates externalities that lead to accelerated growth ([1,2,3]). Yet, lacking formal models, mainstream economic theory has not been able to incorporate these ideas. Instead, two approaches have been used to explain a country's pattern of specialization. The first focuses on the relative proportion between productive factors (i.e. physical capital, labor, land, skills or human capital, infrastructure, and institutions)([4]). Hence, poor countries specialize in goods relatively intensive in unskilled labor and land while richer countries specialize in goods requiring infrastructure, institutions, human and physical capital. The second approach emphasizes technological differences ([5]) and has to be complemented with a theory of what is behind these differences and how they evolve over time. The varieties and quality ladders models ([6,7]) assume a continuum of products, so there is always a slightly more advanced product or just a different one that countries can move to, abstracting away similarities between products when thinking about structural transformation and growth.

Let us assume instead that the underlying productive factors are relatively specific to a set of products. For example, making cotton shirts does not require more or less skills than making chocolate, but different skills. Transporting oil requires pipelines on which you cannot transport fresh fruit, which do require cold storage systems that can be used by other fresh produce as well as institutional setups dealing with phyto-sanitary issues irrelevant for other manufactures. While existing theories abstract from this complexity, here we propose that it underpins the relatedness between products in the fact that countries specializing in one product may or may not also specialize in the other. We empirically study this pattern and use network techniques to show that this relatedness, calculated at a certain time point from trade data, governs how countries change their specialization patterns over time: countries move preferentially to related or 'nearby' goods.

In theory, many possible factors may cause relatedness between products, such as the intensity of broad factors like labor, land, and capital([8]), the level of technological sophistication ([9,10]), the inputs or outputs involved in a product's value chain (e.g. cotton, yarn, cloth, garments) ([11]) or requisite institutions ([12,13]). All of these measures are based on *a priori* notions of what dimension of similarity is most important, assuming that factors of production, technological sophistication or institutional quality exhibit little specificity. Instead, we take an agnostic approach and use an outcomes-based measure, based on the idea that if two goods are related, because they require similar institutions, infrastructure, physical factors, technology, or some combination thereof, then they will tend to be produced in tandem, whereas highly dissimilar goods are less likely to be produced together. Our measure of similarity between products *i* and *j* is based on the conditional probability of having Revealed Comparative Advantage([14]), which measures whether a country is an effective exporter (RCA>1) of a given good *i* or not (RCA<1), given that the country has comparative advantage in good *j* at time *t*, and vice versa. We take the minimum of the pairwise conditional probabilities to have a stringent and symmetric measure,

$$\varphi_{i,j,t} = \min\{P(RCAx_{i,t} \mid RCAx_{j,t}), P(RCAx_{j,t} \mid RCAx_{i,t})\}$$

To calculate these probabilities we use international trade data, cleaned and made compatible([15]) through a National Bureau of Economic Research project lead by Robert Feenstra([16]), disaggregated according to the Standardized International Trade Code at the 4-digit level (SITC-4), providing for each country the value exported to all other countries for 775 products. Using this we calculate the 775 by 775 matrix of revealed proximity between every pair of goods using the equation above.

Figure 1A shows a hierarchically clustered version of the matrix. A smooth and homogeneous product space would imply uniform values (homogenous coloring), while a product-ladder model ([7]) would suggest a matrix with high values (or bright coloring) only along the diagonal. Instead the product space of Fig 1 A appears to be modular ([17]), with some goods highly connected and others disconnected. Furthermore, as a whole the product space is sparse, with $\phi_{ij}$ distributed according to a broad distribution (see Supplementary Material) with 5% of its elements equal to zero, 32% of them smaller than 0.1 and 65% of the entries taking values below 0.2. These significant number of negligible connections calls for a network representation, allowing us to explore the structure of the product space, together with the proximity between products of given classifications and participation in world trade. To offer a representation in which all 775 products are part of a single network we determined the underlying maximum spanning tree (see Supplementary Material) and superposed on it the links with a proximity larger than 0.55 (Fig 1B).

Far from homogenous, the product space appears to have a core-periphery structure (Fig 1B). The core is formed by metal products, machinery and chemicals while the periphery is formed by the rest of the product classes. The products in the top of the periphery belong to fishing, animal, tropical and cereal agriculture. To the left there is a strong peripheral cluster formed by garments and another one belonging to textiles, followed by a second animal agriculture cluster. At the bottom of the network we find a large electronics cluster and to the right of the network we have mining followed by forest and paper products.

The network shows clusters of products somewhat related to the classification introduced by Leamer ([8]), which is based on relative factor intensities. Yet it also introduces a more detailed split of some product classes. For example, machinery is naturally split into two clusters, one consisting of vehicles and heavy machinery, and another one belonging to the electronics industry. The heavy machinery cluster is also strongly interwoven with some products classified as capital intensive such as metal products, but not to similarly classified products such a textiles.

The map obtained can be used to observe the evolution of a country's productive structure. For this purpose we hold the product space fixed and study the dynamics of production within it, although changes in the product space are also an interesting avenue for future research ([18]).

We first explore regional variations in the pattern of specialization for four regions in the product space (Fig 2). Products exported by a region with RCA greater than 1, are marked with black squares. Industrialized countries occupy the core, specialized in a set of closely-related products such as machinery, metal products and chemicals. They also have a considerable participation in more peripheral products such as textiles, forest products and animal agriculture. East-Asian countries have developed RCA in a few distinct clusters along the periphery of the core taking control of the garments, electronics and textile clusters. Latin America and the Caribbean are further out in the periphery and participate in mining, all types of agriculture and the garments sector. Finally sub-Saharan Africa exports very few product types, all of which are in the far periphery of the product space, indicating that each region has a distinguishable pattern of specialization in the product space.

Next, we show that the structure of the product space affects potential changes in a country's pattern of specialization. Figure 3A shows how comparative advantage evolved in Malaysia and Colombia between 1980 and 2000 in the electronics and garments sector respectively. We see that both countries follow a diffusion process in which comparative advantage moves preferentially towards products close to existing goods: garments in Colombia and electronics in Malaysia.

Beyond this graphical illustration, is it really true that countries develop comparative advantage preferentially in nearby goods? We use two different approaches to answer this question. We first develop a measure of the average proximity of a new potential product $j$ to a country's current productive capability, which we call *density* and define as

$$\omega_j^k = \frac{\sum_i x_i \phi_{ij}}{\sum_i \phi_{ij}},$$

where $\omega_j^k$ is the density around good $j$ for the $k^{th}$ country, $x_i = 1$ if $RCA_i > 1$ and 0 otherwise and $\varphi_{ij}$ is the matrix of proximities between goods $i$ and $j$. A value of $\omega_j$ equal to 0.4 for a given product/country means that in a particular country from the perspective of the $j^{th}$ product, 40 percent of the neighboring space seems to be developed. To study the evolution in the patterns of comparative advantage we consider products exported with an RCA less than 0.5 in 1990 and look at the same products in 1995. For each country, we call *transition products* those exported with RCA>1 in 1995 and *undeveloped products* as those that in 1995 had an RCA< 0.5, disregarding the rest as inconclusive. Figure 3B shows the distribution of density around transition products (yellow) and compares it to densities around undeveloped products (red). Clearly, these distributions are very distinct, with a higher density around transition products than among undeveloped ones.

We deepen this approach by considering the density around each product in countries that transitioned into them and compare it to the density in the countries that did not develop comparative advantage in the product,

$$H_j = \frac{\sum_{k=1}^{T} \omega_j^k / T}{\sum_{k=T+1}^{N} \omega_j^k / (N-T)}$$

where we average the $\omega_j$ across the $T$ countries that transitioned into the good and the ($N-T$) countries that did not. Figure 3C shows the frequency distribution of this ratio. For 79 percent of products, this ratio is greater than 1 indicating that the density in countries that transitioned is greater than in countries that did not, often substantially.

An alternative way of illustrating that countries develop RCA in goods close to the ones where they had, is to calculate the conditional probability of transitioning into a product if the nearest product with RCA>1 is at a proximity φ. Figure 3D shows a monotonic relationship between proximity of the nearest good and the probability of transitioning into it. While the probability of moving into a good at proximity 0.1 in the course of 5 years is almost nil, the probability is about 15 percent if the closest good is at proximity of 0.8 ([19]).

Since production shifts to nearby products, we ask whether the product space is sufficiently connected that given sufficient time, all countries can reach most of it, particularly its dense core. Lack of connectedness may explain the difficulties faced by countries trying to converge to the income levels of rich countries: they may not be able to undergo structural transformation because proximities are too low. A simple approach is to calculate the relative size of the largest connected component as a function of φ. Figure 3E shows that at $\phi \geq 0.6$ the largest connected component has a negligible size compared with the total number of products while for $\phi \leq 0.3$ the product space is almost fully connected, meaning that there is almost always a path between two different products.

We study the impact of the product space structure by simulating how the position of countries evolve if they were allowed to repeatedly move to all products with proximities greater than a certain $\phi_o$. If countries are able to diffuse only to nearby products, but these are sufficiently connected to others, then after several iterations, 20 in our exercise, countries would be able to reach richer parts of the product space. On the other hand, if the product space is sufficiently disconnected and modular, countries will not be able to find paths to the richer part of the product space, independently of how many iterations they are allowed to make at proximities above $\phi_o$.

The results of our simulation for Chile and Korea are presented in Figure 4A. At a relatively low proximity ($\phi$=0.55) both countries are able to diffuse through to the core of the product space, however Korea is able to do so much faster thanks to its current presence in core products. For higher proximities the question becomes whether the country can spread at all. Diffusing up to a proximity of $\phi$=0.6 Chile is able to spread very slowly throughout the space while Korea is still able to populate the core after 4 rounds. At even higher proximities, $\phi=0.65$, Chile is not able to diffuse at all and Korea develops RCA slowly to a few products close to the machinery and electronics cluster.

To generalize this analysis for the whole world, we need a summary measure of the position of a country in the product space. We adopt a measure based on Hausmann, Hwang and Rodrik ([20]), which involves a two stage process. First, for every product we assign an income level, which is the weighted GDP per capita of countries with comparative advantage in that good, which we call PRODY ([20]). We then choose the average of the PRODYs of the top $N$ products that the country has access to. We characterize the position of a country in the product space after $M$ iterations at proximities above $\varphi_o$ as the average PRODY of the top $N$ products that can be reached and denote it by $<PRODY>_{M\phi}^{N}$. We show the distribution of $<PRODY>_{M\phi}^{N}$ in Figure 4B for $N = 50$, $M=20$ and $\phi=1$ (green), $\phi=0.65$ (yellow) and $\phi=0.55$ (red). The distribution for $\phi=1$ allows us to characterize the current distribution of countries in the product space. It shows a bimodal distribution, signature of a world divided into rich and poor countries with few countries occupying the center of the distribution. When we allow countries to diffuse up to $\phi=0.65$, this distribution does not change significantly: it shifts slightly to the right due to the acquisition of a limited number of sophisticated products by some countries. This diffusion process, however, stops after a few rounds and the world maintains a degree of inequality similar to its current state. Contrarily, when the diffusion parameter is lowered to $\phi=0.55$, most countries are able to diffuse and reach the most sophisticated basket in the long run. Only a few countries are left behind, which unsurprisingly make up the poorest end of the income distribution.

To quantify the level of convergence we calculated the Inter Quartile Range (IQR) for the distribution of $<PRODY>_{20\phi}^{50}$ and normalize this quantity by dividing it with the IQR for the original distribution. Figure 4C shows that the level of convergence of the system goes through an abrupt transition and that convergence is possible if countries are able to diffuse to products located at a proximity $\phi>0.65$.

The bi-modal distribution of international income levels and a lack of convergence of the poor towards the rich has been explained using geographic ([21]) and institutional ([12,13]) arguments. Here, we introduced a new factor into this discussion: the difficulties involved in moving through the product space. The detailed structure of the product space is shown here for the very first time and together with the location of the countries and the characteristics of the diffusion process undergone by them, strongly suggests that not all countries face the same opportunities when it comes to development. Poorer countries tend to be located in the periphery of the product space in which the move towards new products is harder to achieve. More interestingly, among countries with a similar level of development and seemingly similar levels of production and export sophistication, there is significant variation in the option set implied by their current productive structure, with some on a path to continued structural transformation and growth, while others are stuck in a dead end.

These findings have important consequences for economic policy, as the incentives to promote structural transformation in the presence of proximate opportunities are likely quite different from that required when a country hits a dead-end. It is quite difficult for production to shift to far-away products in the space, and therefore policies to promote large jumps are more challenging. Yet, precisely these long jumps are the ones generating new options for subsequent structural transformation.

In this paper we have mapped the product space, and studied how countries dynamically evolve on it. We empirically proved that countries develop RCA following a diffusion process for which our outcome based definition of the product space appears to be the natural substrate. Moreover, the structure of the product space limits the diffusion process by being non-traversable by jumps of any proximity. When we combine these results with the fact that poorer countries tend to have RCA mainly on peripheral products, it implies that a country's productive structure is constrained not only by its levels of factor endowments, but also by how easily those product-specific factors can be adapted to alternative uses, as indicated by location in the product space. On a more global perspective, these results point towards a new hypothesis for the lack of income convergence in the world: convergence can only exist if countries have the ability to reach any area of the product space. Our study shows that most of the diffusion occurs through links with proximities of 0.6 or larger, thus the most popular strategy involves diffusing to nearby products, a strategy that is successful for richer countries located on the core of the space, and ineffective for poorer countries populating the periphery.

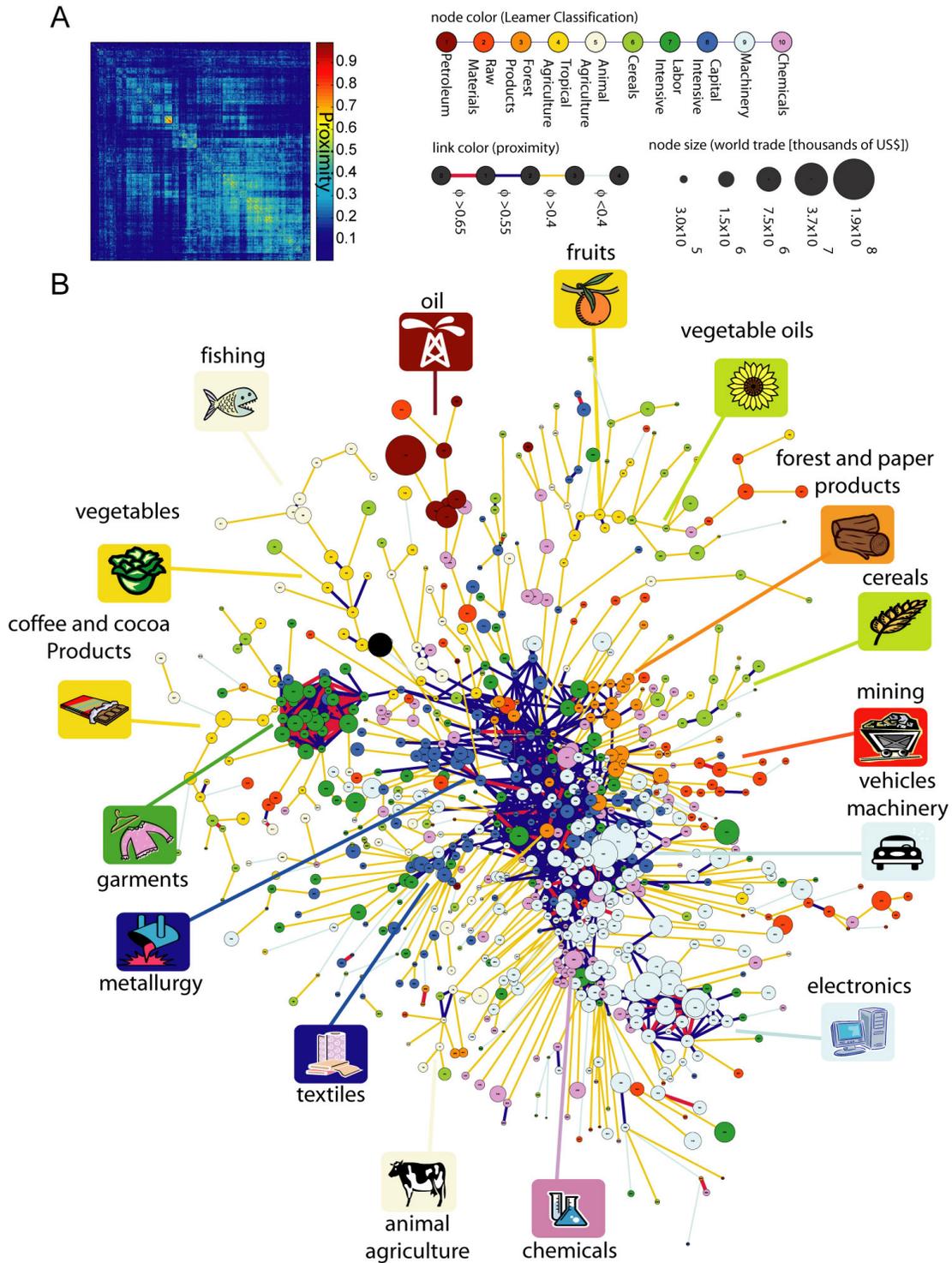

*Figure 1. The product space and Leamer clusters. **A.** Hierarchically clustered proximity matrix representing the 775 SITC-4 product classes exported in the 1998-2000 period. **B.** Network representation of the product space. This network was laid out using a force spring algorithm and retouched by hand .*

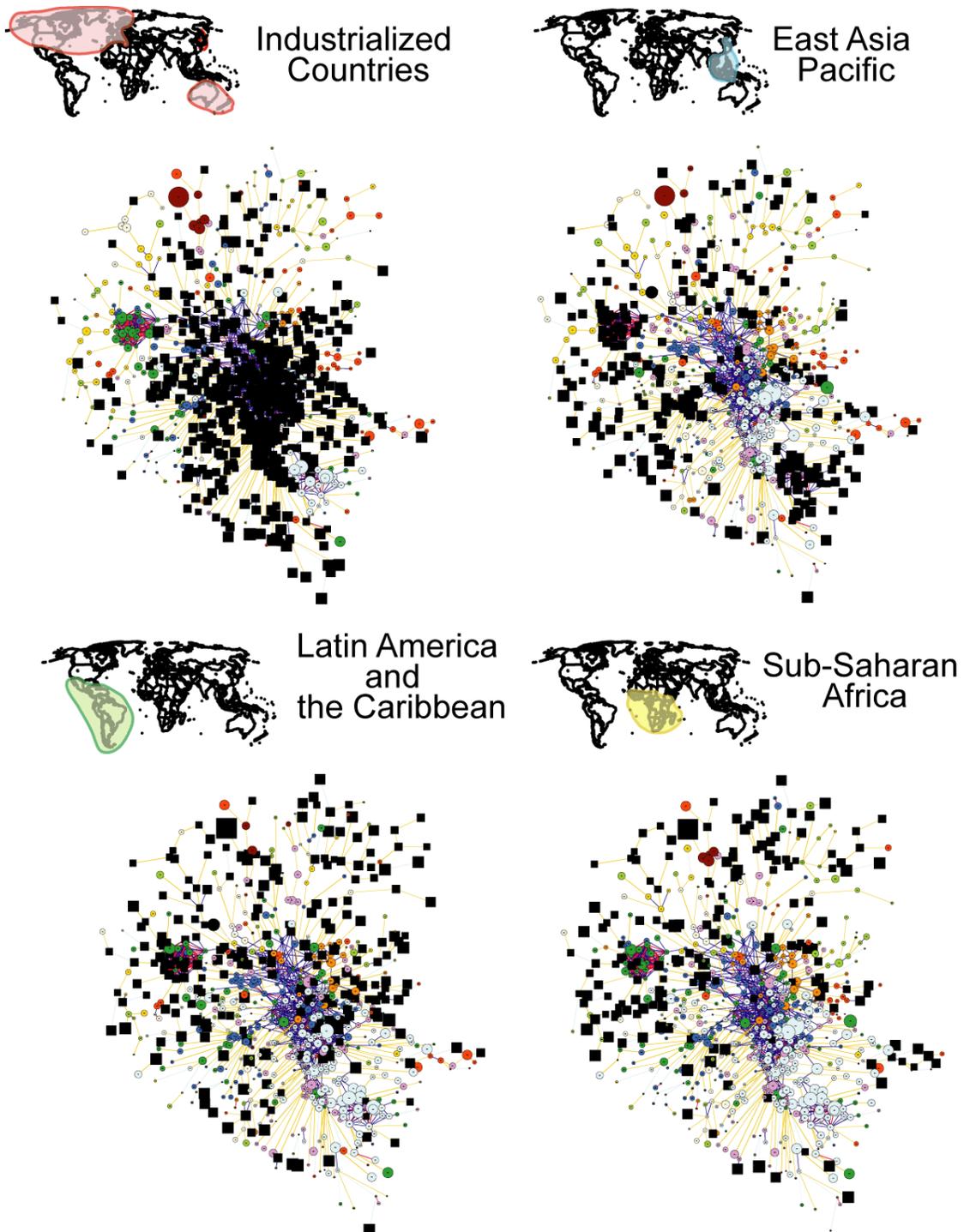

*Figure 2. Localization of the productive structure for different regions of the world. The products for which the region has an RCA > 1 are denoted by black squares.*

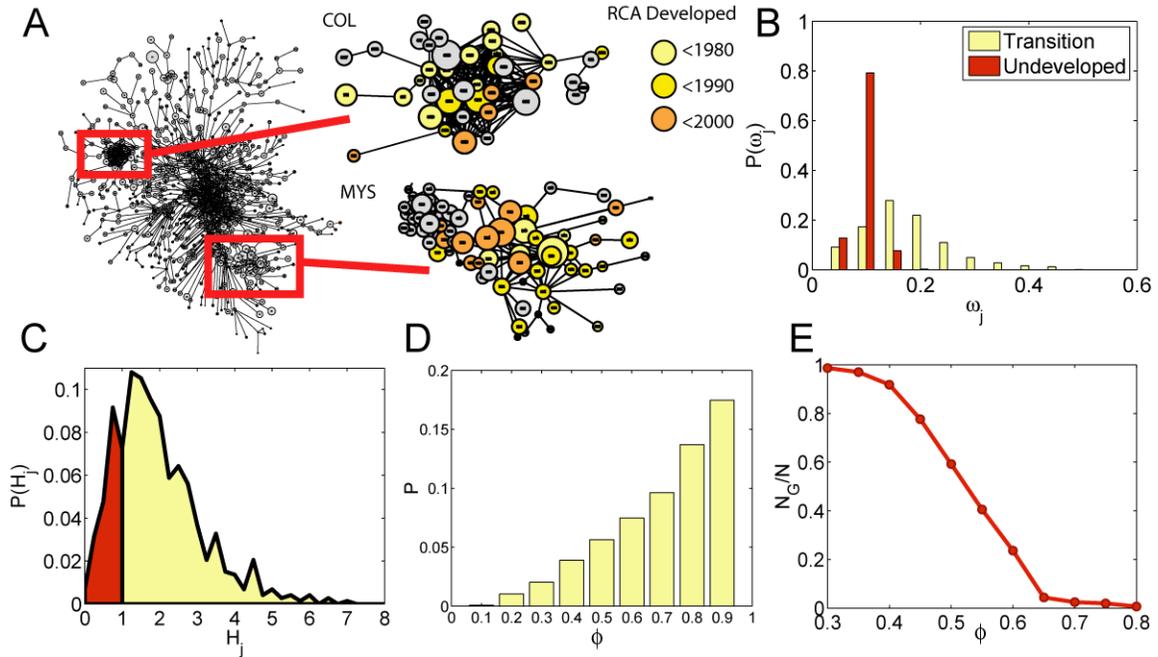

*Figure 3*. *Empirical evolution of countries*. ***A.*** *Examples of RCA spreading for Colombia (COL) and Malaysia (MYS). The color code shows when this countries first developed RCA>1 for products in the garments sector in Colombia and the electronics cluster for Malaysia.* ***B.*** *Distribution of density for transition products and undeveloped products* ***C.*** *Distribution for the relative increase in density for products undergoing a transition with respect to the same products when they remain undeveloped.* ***D.*** *Probability of developing RCA given that the closest connected product is at proximity* $\phi$. ***E.*** *Relative size the largest connected component $N_G$ with respect to the total number of products in the system N as a function of proximity* $\phi$.

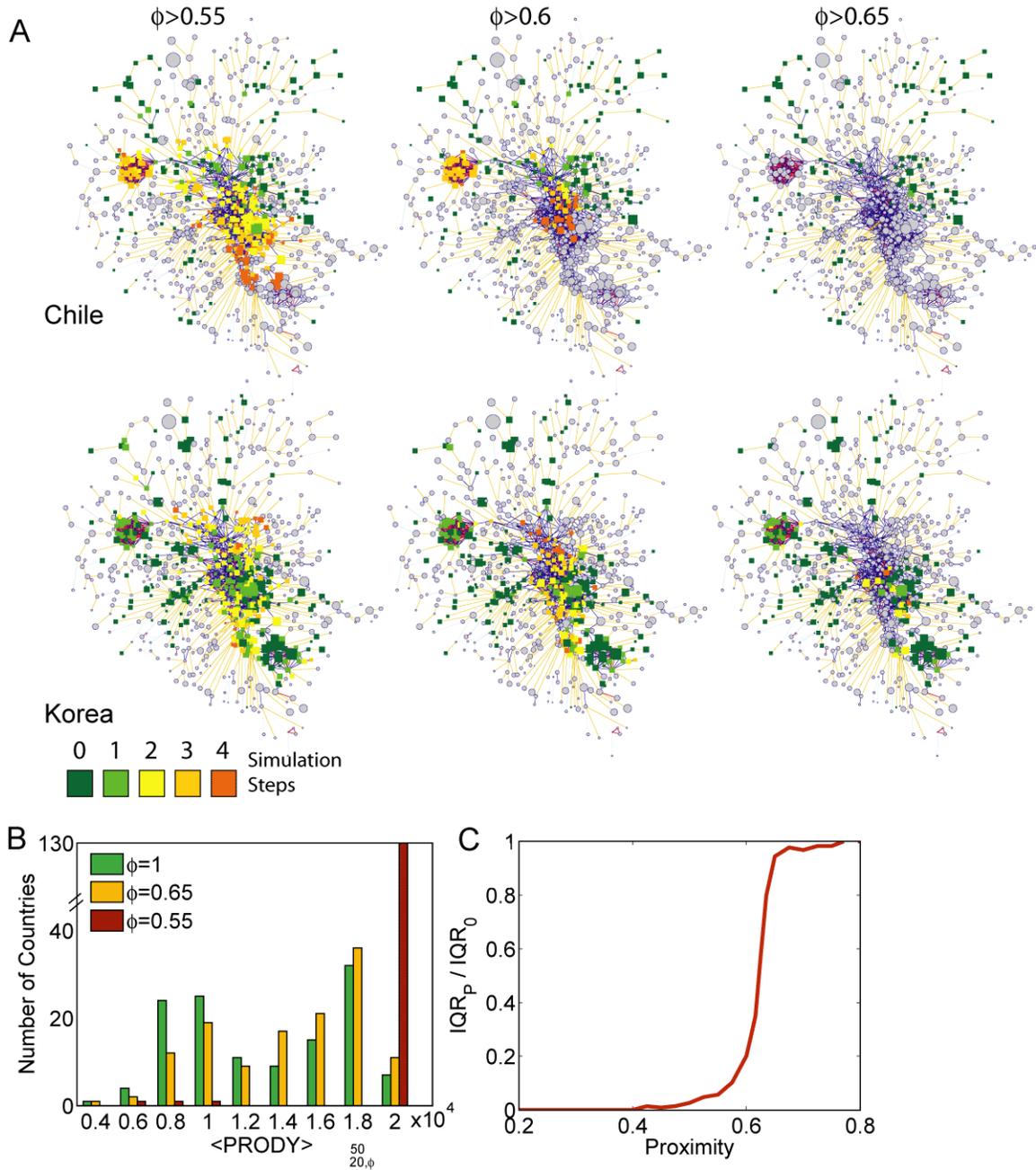

*Figure 4. Simulated diffusion process and inequality. **A.** Simulated diffusion process for Chile and Korea in which we allow countries to develop RCA in all products closer than ϕ= 0.55, 0.6 and 0.65. The number of steps required to develop RCA can be read from the color code on the top right corner of the figure. **B.** Distribution for the average PRODY of the best 50 products in a countries basket before and after 20 rounds of diffusion. The original distribution is shown in green while the one associated with the distribution after 20 diffusion rounds with ϕ=0.65 is presented in yellow and ϕ=0.55 in red. **C.** Global convergence of the diffusion process as a function of proximity measured by calculating the inter quartile range after diffusing with a given ϕ normalized by the inter quartile range of the original distribution.*

the product space, where transitions are also less frequent. Thus, densely connected products can develop RCA through more paths than sparsely connected ones, indicating the importance of absolute proximity
20. We follow the methodology developed in Hausmann, Hwang and Rodrik (2006), which weighs the GDP per capita of each country exporting that product by the RCA that the country has in that good. R. Hausmann, J. Hwang, D. Rodrik, *What you export matters* (NBER Working Paper 11905, Cambridge MA 2006)
21. J. Gallup, J. Sachs, A. Mellinger, Geography and Economic Development Gallup et al. International Regional Science Review.1999; 22: 179-232